\input{psfig}
\def\puncspace{\ifmmode\,\else{\ifcat.\C{\if.\C\else\if,\C\else\if?\C\else%
\if:\C\else\if;\C\else\if-\C\else\if)\C\else\if/\C\else\if]\C\else\if'\C%
\else\space\fi\fi\fi\fi\fi\fi\fi\fi\fi\fi}%
\else\if\empty\C\else\if\space\C\else\space\fi\fi\fi}\fi}
\def\SP{\let\\=\empty\futurelet\C\puncspace}
\def\kms{km~s$^{-1}$}
\def\etal{{\it et al.\/}\ }

\def\eg{{\it e.g.\/}\rm,\ }
\def\void#1{{}}
\def\lsim{~\rlap{$<$}{\lower 1.0ex\hbox{$\sim$}}}
\def\gsim{~\rlap{$>$}{\lower 1.0ex\hbox{$\sim$}}}

\documentstyle[11pt,paspconf]{article}

\begin{document}

\title{The Nearby Early-type Galaxies Survey (ENEAR): Project Description and 
Some Preliminary Results}

\author{Gary Wegner}
\affil{Department of Physics \& Astronomy, Dartmouth College, Hanover, NH 
03755-3528, U. S. A.}

\author{L. N. da Costa}
\affil{European Southern Observatory, Karl-Schwarzschild Stra$\ss$e 2, D-85748 
Garching bei M\"unchen, Germany and Departamento de Astronomia,
Observat$\acute{o}$rio Nacional, Rua General Jos$\acute{e}$ Cristino
77, Rio de Janiero, R. J. 20921 Brazil}

\author{M. V. Alonso}
\affil{Observatorio Astr$\acute{o}$nomico de Cord$\acute{o}$ba Laprida 
854, Cord$\acute{o}$ba (5000), Argentina}

\author{M. Bernardi}
\affil{European Southern Observatory, Karl-Schwarzschild Stra$\ss$e 2, D-85748 
Garching bei M\"unchen and Universit\"ats-Sternwarte M\"unchen, 
Scheinerstra$\ss$e 1, D-81679, Germany}

\author{C. N. A. Wilmer}
\affil{ Departamento de Astronomia, Observat$\acute{o}$rio Nacional, Rua General 
Jos$\acute{e}$ 
Cristino 77, Rio de Janiero, R. J. 20921 Brazil and
UCO/Lick Observatory, University of California, 1156 High Street, Santa Cruz, 
CA 95064, U.S.A.}

\author{P. S. Pellegrini, C. Rit\'e, M. Maia}
\affil{Departamento de Astronomia,  Observat$\acute{o}$rio Nacional, Rua General 
Jos$\acute{e}$ Cristino 77, Rio de Janiero, R. J. 20921 Brazil}

\begin{abstract}

The ENEAR project is an all-sky survey of nearby early-type galaxies.
About 2200 new spectra and $R$-band images of over 1500 galaxies
together with published data were assembled into a large catalog
containing over 2000 objects with $cz$, $\sigma$, photometric data, and
line strengths on a uniform system. From this extensive database a
magnitude-limited sample has been drawn comprising $\sim$ 1400 galaxies
brighter than $m_B = 14.5$, $cz < 7000$ \kms, and $T \leq -2$ with
measured distances (ENEARf) and about 500 early-type galaxies in 28
clusters/groups (ENEARc) to derive an internally consistent $D_n-\sigma$
relation to estimate galaxy distances. In this contribution we discuss
some general properties of ENEAR and briefly describe some preliminary
results. 

\end{abstract}


\section{Introduction}
The importance of cosmic flows to cosmology and to our understanding of
the origin and evolution of large-scale structures as well as  earlier
observational and theoretical work on the subject have been reviewed by
Strauss (1999) and Frenk (1999) in this workshop.  On the observational
side attention has focused on using promising new accurate distance
indicators, such as the SBF method (Tonry \etal 1999)
and SNeIa (Schmidt 1999, Riess 1999). Nevertheless, the FP and TF
methods have remained the dominant distance indicators in this field
because so far the new methods either do not reach great enough depths
or do not provide enough objects for a detailed mapping of the peculiar
velocity field.

Most of the recent work in the field involving all-sky samples such as
the SFI/SCI (da Costa \etal 1996 and Giovanelli \etal 1997), Mark III
(Willick \etal 1997), and SHELLFLOW (Courteau \etal 1999) have relied
primarily on TF samples. Other recent investigations, using different
distance indicators, have been specialized in one way or another (\eg
EFAR Colless \etal 1999, Wegner 1996, SMAC Hudson \etal 1999, and LP10K
Willick \etal 1999) and with few exceptions have concentrated on
pre-chosen areas of the sky which do not render them optimal for all
types of analyses.  However, it is apparent from theoretical studies and
discussions in this workshop (\eg Davis 1999, Burstein 1999, Hudson
1999, Kolatt \etal 1999, Zehavi 1999, Hoffman 1999) that there is much
information to be gained from large dense surveys of peculiar velocities
where selection and errors are well understood. 
   
Developments in instrumentation, particularly CCDs, have greatly aided
in obtaining improved spectra and photometry and brought much larger
samples of improved data quality within reach. The new redshift-distance
survey of a magnitude-limited sample of early-type galaxies  reported
here, comprising 1400 galaxies in about 900 objects within 7000 \kms, is
a substantial enlargement of the available data.  The last all-sky
elliptical galaxy survey, containing about 550 galaxies in 226 objects
is still the 7S study (Faber \etal 1989). 

For the past several years, the authors combined their resources to
carry out an all-sky survey of nearby early-type field and cluster
galaxies (E and S0) in order to utilize the FP relation to measure their
$V_{pec}$ for redshifts $cz < 7000$ \kms, hence the name `ENEAR.' The
general purpose of ENEAR was to study large-scale motions based  on
early-type galaxies to a comparable depth to that reached by the SFI TF
sample.  Such a sample will allow a comparison 
between the motions of spirals and
ellipticals, based on different distance relations, and their
combination  will provide an even denser set of probes of gravity in the
nearby universe. In addition, the new spectroscopy provides the
important by-product of linestrengths with which galaxies in different
environments can be compared.

\section{Definition of the ENEARf Sample}

The ENEARf sample provides all-sky coverage, except for the usual lack
of data along the galactic equator, and consists of galaxies with
$m_B$ brighter than 14.5 mag and $cz < 7000$ \kms. Morphological types
are limited to $T \leq -2$.  A number of catalogs were used to
construct the initial all-sky catalog from which the early-type
subsample was drawn. Among them CfA1 (Huchra \etal 1983), SSRS (da
Costa \etal 1991) and the equatorial survey (Huchra \etal 1990) at
high galactic latitudes (Pellegrini \etal 1990), and the ORS sample
(Santiago \etal 1995) at lower $|b|$. Combined these samples give full
sky coverage to the desired depth, but required slight adjustments in
magnitude and morphological type scales. A more detailed description
of the sample will be presented elsewhere (da Costa \etal 2000a).

\section{The New Observations}

The new observations consist of 2200 spectra of about 1650 galaxies and
2100 $R$-band CCD images of 1580 galaxies. The bulk of the northern
hemisphere observations were made at the MDM Observatory using the 2.4 m
Hiltner and 1.3 m McGraw-Hill telescopes. The southern hemisphere
imaging observations were made using different telescopes at ESO and 
the 0.9m at CTIO, while the spectroscopic observations were carried out
primarily at the ESO 1.5m telescope. The data from both hemispheres were
overlapped in order to get a good handle on possible systematic errors.
Northern hemisphere data extended to $\delta = -20\deg$ and southern
hemisphere observations were carried out to $\delta = +20\deg$.

The observational setups, procedures, and data reductions are similar to
those given in Wegner \etal (1999) and Saglia \etal (1997). They will be
described in detail in Wegner \etal (2000) and Alonso \etal (2000). For
spectroscopic data, there are about 500 galaxies with two or more
observations and about 200 galaxies with measurements available in the
literature. We estimate that our internal errors are typically 8\% in
velocity dispersion and 0.01~mag in Mg$_2$. For photometric data, there
are about 500 galaxies with repeated observations and over 300 with
measurements publicly available. Errors in the global parameters are
estimated to be: 0.017~dex in $\log{D_n}$, 0.08~dex in $\log{r_e}$,
0.3~mag/arcsec$^2$ in $\bar{\mu}_e$, 0.019 in  FP = $\log{r_e}$ - 0.30
$\bar{\mu}_e$, and 0.09~mag in the total magnitude. The large number of
repeated observations and the overlap with published data sets is key to
bring all measurements into a common a system and to produce a
homogeneous sample.

\section{Sky Coverage and Completeness}

\begin{figure}
\centering
\mbox{\psfig{figure=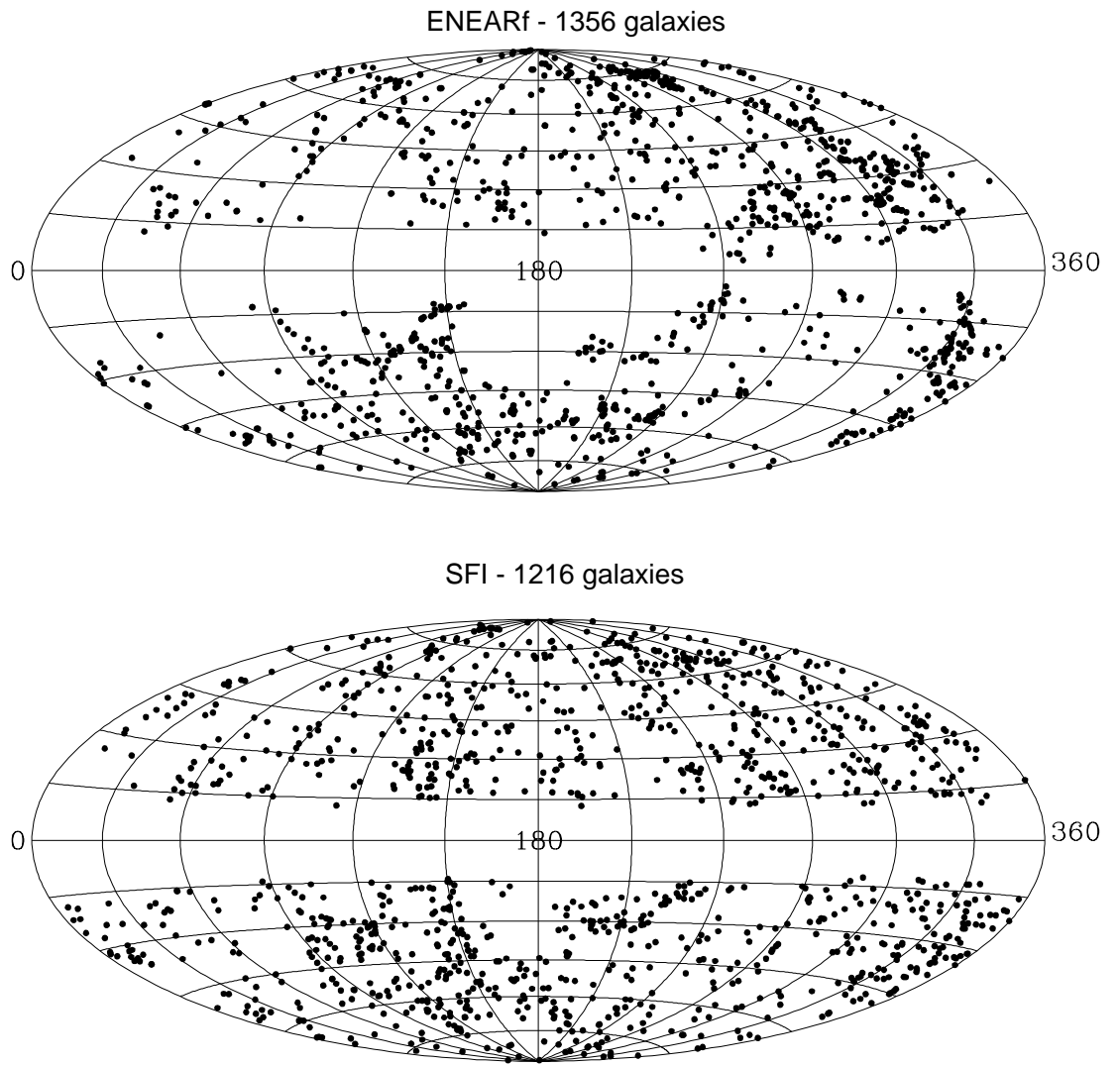,height=16truecm,bbllx=4truecm,bblly=5truecm,bburx=19truecm,bbury=25truecm}}
\caption{Comparison of the projected distributions of galaxies with
available peculiar velocities on the sky in galactic coordinates for
the ENEARf and the SFI samples.}
\label{fig-1}
\end{figure}

\begin{figure}
\centering
\mbox{\psfig{figure=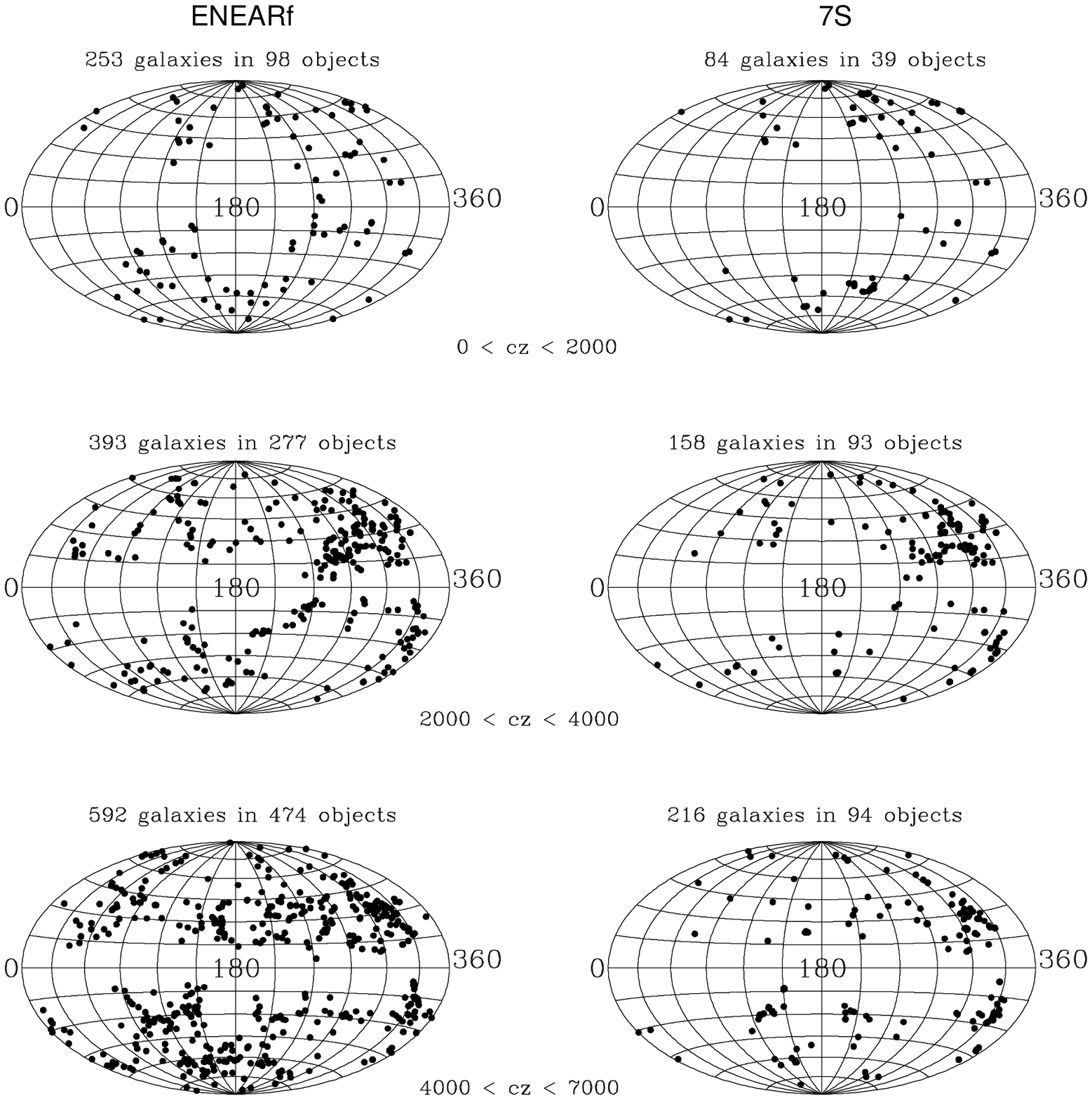,height=16truecm,bbllx=4truecm,bblly=5truecm,bburx=19truecm,bbury=25truecm}}
\caption{Comparison of the early-type galaxies observed in the ENEARf
sample and the 7S survey projected on the sky in galactic 
coordinates and in three redshift shells.} \label{fig-2}
\end{figure}

The projected positions of the observed ENEARf galaxies on the sky in
galactic coordinates is shown in Figure~\ref{fig-1} where they are
compared with the distribution of the spiral galaxies from the SFI
sample. As can be seen while the total number of galaxies with
measured distances is comparable to that of SFI, the distribution of
ENEAR galaxies is visibly more clumpy, better delineating the
high-density regions of the nearby universe. After assigning galaxies
to groups the total number of independent objects is $\sim$
850. Figure~\ref{fig-2} compares the ENEARf sample with that of the 7S
(E's) in three redshift shells, after grouping. This figure
illustrates the relatively dense sampling accomplished by the ENEAR
project compared to the older data. It is important to note the
differences between the two samples in the Perseus-Pisces region, a
source of great debate in the past.

The completeness of the sample, as judged by a comparison with the total
number of objects that could have been observed, is about 70\% complete
within the chosen redshift and magnitude cutoff. The remaining galaxies
are predominatly lenticular galaxies. However, it is important to
emphasize  that the level of completeness of the sample is uniform
across the sky.

\section{Groups and Clusters}

The ENEAR sample also complements recent work on early-type galaxies in
clusters; at the moment 446 elliptical and S0 galaxies have been
assigned to 28 clusters (ENEARc) and used to derive the $D_n-\sigma$
relation. Some of the clusters in the ENEARc sample include ENEARf
galaxies while others lie beyond 7000\kms to include previously
well-studied clusters. Information on the cluster sample and the
distances and motions of the clusters will be reported in Bernardi \etal
(2000a, b).

In contrast to earlier work, groups and clusters with $cz \lsim
7000$~\kms were identified by an objective algorithm applied to the
original all-sky sample with complete redshift information. Fainter
galaxies were assigned to the identified groups or clusters using
well-defined criteria. Systems with more than five early-type member
galaxies were used in the definition of the distance relation. The
adopted membership assignment procedure is considerably superior to
those utilized in the past when only incomplete redshift information was
available.  The identification of groups also permits a much improved
grouping of galaxies, which is critical for studies of the peculiar
velocity field using early-type galaxies. 

In Bernardi \etal (1998) we have used this objective membership
assignment to compare the properties of early-type galaxies in different
environments using the Mg$_2$ line index and the central velocity
dispersion $\sigma$. We have found that the Mg$_2-\sigma$ relation does
not show any dependence on environment. This result favors a scenario of
formation and evolution of early-type galaxies in which the bulk of
stellar populations in galactic spheroids formed at high redshift
($z\gsim 3$), no matter whether such spheroids now reside in high or low
density regions. Furthermore, the lack of environmental effects also
supports the assumption that the empirical distance indicators such as
the $D_n-\sigma$ relation determined using cluster galaxies, are
universal; this is a crucial assumption for peculiar velocity field
analyses.

\section{Distances and Motions}

We use our homogeneous data for 28 clusters/groups to determine an
internally consistent, bias corrected $D_n-\sigma$ relation (Bernardi
\etal 2000b). We have fitted the following regression: $$\log D_n =
1.180 \log \sigma + 1.391.$$  These coefficients are quite close to
those found by the 7S (Faber \etal 1989). The scatter in the derived
distance relation corresponds to a distance error of about 19\% per
galaxy, comparable to what is obtained for the FP relation.  The above
$D_n-\sigma$ relation is used to find distances and peculiar velocities
for all objects (individual galaxies and groups) in the ENEARf sample.
Using the radial components of the peculiar velocity we fit a a dipole
flow model for which we find  (da Costa \etal 2000b):

$V = 343 \pm 53$ ~\kms,~
$l = 264\deg \pm 8\deg$, and~
$b = 33\deg \pm 5\deg$.

This direction is close to the direction of the LG motion and consistent
to the one found in the recent cluster study of Dale  \etal (1999).

\section{Conclusions}

The observational phase of the ENEAR project is approaching completion.
More detailed descriptions of all portions of the project and analysis
will be given in a forthcoming series of papers. At this time our main
goal has been to show the scope and coverage of the database. For
peculiar motions, the ENEAR sample complements the SFI study for spirals
in depth, number of galaxies, and accuracy of observations. Our
preliminary look at the data shows that the observed peculiar motions of
the early-type galaxies in ENEAR  is similar to that found
from the SFI (Zaroubi \etal 2000), suggesting that an accurate picture
of the velocity field in the local universe is finally emerging.

\acknowledgements  We thank the collaboration of S. Zaroubi in the
preliminary analysis of the data. We are also grateful to Ot\'avio
Chaves for his enormous contribution in different phases of this
project. The spectroscopic observations in the southern hemisphere were
conducted at the ESO 1.5m under the agreement between ESO and the
Observat\'orio Nacional.

\end{document}